\documentclass[a4paper,11pt]{article}
\usepackage{mathrsfs}
\usepackage{amsmath,amsfonts,amssymb,amsthm, bm}
\usepackage{graphicx}
\usepackage[caption = false]{subfig}
\usepackage[normalem]{ulem}
\usepackage{xcolor}
\definecolor{lcolor}{rgb}{0.5,0,0}
\definecolor{citcolor}{rgb}{0,0.3,0.0}
\usepackage{nicefrac}
\usepackage{comment}
\usepackage{tabularx}

\usepackage{mciteplus}

\usepackage{physics}

\usepackage{enumitem}
\usepackage{pos}

\newcommand{\xpom}{{x_\mathbb{P}}}
\newcommand{\xit}{{\boldsymbol{\xi}}}
\newcommand{\xt}{{\mathbf{x}}}
\newcommand{\ud}{\, \mathrm{d}}
\newcommand{\as}{\alpha_{\mathrm{s}}}
\newcommand{\zt}{{\mathbf{z}}}
\newcommand{\Kt}{{\mathbf{K}}}

\title{Diffractive structure functions from JIMWLK evolution}

\author{Tuomas Lappi}
\author{Heikki Mäntysaari}
\author*{Pyry Runko}

\affiliation{
Department of Physics, University of Jyväskylä,\\  P.O. Box 35, 40014 University of Jyväskylä, Finland
}
\affiliation{
Helsinki Institute of Physics,\\ P.O. Box 64, 00014 University of Helsinki, Finland
}

\emailAdd{tuomas.v.v.lappi@jyu.fi}
\emailAdd{heikki.mantysaari@jyu.fi}

\emailAdd{pyry.j.runko@jyu.fi}
\abstract{We compute diffractive structure functions from Wilson line configurations whose energy evolution is given by the JIMWLK equation. We use a JIMWLK evolution setup that has already been constrained with exclusive vector meson production data from HERA. We compare our results to HERA measurements and also extended to heavy nuclei. In particular we can calculate predictions for the nuclear modification factor and diffractive-to-total cross section ratios at the EIC.}

\FullConference{The 33rd International Workshop on Deep Inelastic Scattering and Related Subjects (DIS2026)\\
4 - 8 May 2026\\
Bologna, Italy\\}

\begin{document}
\maketitle

\section{Introduction}
The goal of this work is to predict the $\xpom$ dependence of diffractive structure functions using the JIMWLK evolution~\cite{Mueller:2001uk}. We compare our results for the proton to the combined HERA reduced cross section data~\cite{H1:2012xlc} to establish a baseline. The main result of this work is the nuclear modification of the diffractive structure functions for the gold nucleus in Electron-Ion Collider (EIC) kinematics~\cite{Accardi:2012qut}.

\begin{figure}[h]
  \begin{tabular}{ll}
  \begin{minipage}[t]{0.5\textwidth} 
  \centering
    \includegraphics[width=\linewidth]{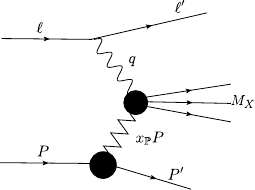}\\
     \textbf{(a)} DDIS
    \end{minipage}
    &
  \begin{minipage}[t]{0.4\textwidth} 
  \centering
    \includegraphics[width=\linewidth]{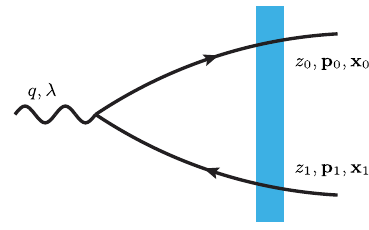}\\
    \textbf{(b)}  Dipole picture amplitude
    \end{minipage}
  \end{tabular}
  \caption{\textbf{(a)} Diagram depicting DDIS $\ell + A \to \ell +A +X$. A charged lepton $\ell$ scatters with a nucleus via a virtual photon with four momentum $q$ and a pomeron carrying a fraction $\xpom$ of the target momentum $P$, producing additional particles in the final state with invariant mass $M_X$. \textbf{(b)} Dipole picture scattering amplitude diagram. A virtual photon with polarization $\lambda$ and four-momentum $q$ splits into a quark-antiquark ($q\bar q$) pair with momentum fractions $z_0$ and $z_1=1-z_0$. The $q\bar q$ dipole interaction with the color field of the target nucleus is depicted here by the shaded band.}
    \label{fig:DDIS&dipole}
\end{figure}

In diffractive deep inelastic scattering (DDIS), depicted in Fig.~\ref{fig:DDIS&dipole}a, a lepton scatters with a nucleus and produces a diffractive system $X$ in the final state without exchanging net color charge.  We consider here only coherent scattering where the nucleus stays intact. Experimentally, diffractive events are characterized by a particle-free region in the detector between the outgoing nucleus and produced particles $X$. This is called a rapidity gap.

The relevant kinematical variables of DDIS are the virtuality of the photon $Q^2=-q^2$ defined by its four momentum $q$, the variable $\beta \approx \frac{Q^2}{M^2_X+Q^2}$, which controls the invariant mass $M_X$ of the diffractive system, and $\xpom$, the fraction of the target momentum  that is carried by the pomeron. The variable $\beta$ can also be interpreted as the fraction of the momentum $\xpom P$ carried by a parton inside the pomeron that interacts with the virtual photon.

Diffractive structure functions can be computed from the diffractive virtual photon-nucleus cross section
\begin{equation}
    \xpom F_\lambda^{\text D(4)} = \frac{Q^2}{4\pi^2 \alpha_{\text{em}}}\frac{Q^2}{\beta}\frac{\text d \sigma_{\gamma^*_\lambda+A}^{\text D}}{\text d M^2_X\text d t},
\end{equation}
where $\alpha_{\text{em}}$ is the electromagnetic fine structure constant, $\lambda$ is the polarization of the virtual photon, either transverse or longitudinal, and $t$ is the Mandelstam variable. The superscript $(4)$ denotes the dependence on the four kinematical variables $\beta$, $Q^2$, $\xpom$ and $t$.

At leading order in the dipole picture the virtual photon fluctuates into a quark-antiquark ($q\bar q$) pair that interacts with the color fields of the target as shown in Fig.~\ref{fig:DDIS&dipole}b. The higher order Fock states with additional gluons interacting with the target, such as $q\bar qg$, are suppressed in the kinematic region of $\beta > 0.5$ which we consider here. In the color glass condensate (CGC) framework the target color fields are classical and described using Wilson lines that are path ordered integrals of the color fields.

A general form for the diffractive $q\bar q$ production cross section is~\cite{Beuf:2022kyp}
\begin{equation}
\label{eq:gammaAxsection}
    \begin{aligned}
        \sigma_{\gamma_\lambda^*\to q\bar q}^\mathrm{D} = &\int \frac{\text d^2\mathbf{p}_0\text d p_0^+\text d^2\mathbf{p}_1\text d p_1^+}{2p_0^+ (2\pi)^3 2p_1^+ (2\pi)^3}2q^+ 2\pi\delta(p_0^++p_1^+-q^+)N_{\mathrm{c}}\\
        &\int \mathrm{d}^2 \mathbf{x}_0 \int \mathrm{d}^2 \mathbf{x}_1 \int \mathrm{d}^2 \bar{ \mathbf{x}}_0 \int \mathrm{d} \bar{\mathbf{x}}_1
        e^{-i\mathbf{x_{0\bar0}(\mathbf{p}_0}-z_0\mathbf{q})}e^{-i\mathbf{x_{1\bar1}(\mathbf{p}_1}-z_1\mathbf{q})}\\
        &\sum_{f,h_0,h_1} \Big(\Tilde{\psi}_{\gamma^*_\lambda \to q_{\bar 0}\bar q_{\bar 1}}\Big)^\dagger
        \Big(\Tilde{\psi}_{\gamma^*_\lambda \to q_{0}\bar q_{1}}\Big)
        \Big[S^\dagger_{\bar 0\bar 1}-1\Big]\Big[S_{01}-1\Big]
    \end{aligned}
\end{equation}
where the dipole scattering amplitude
\begin{equation}
    S_{01}= \frac{1}{N_{\mathrm{c}}}\text{Tr}\big[ V(\mathbf{x}_0)V^\dagger(\mathbf{x}_1)\big] 
\end{equation}
is a trace of the Wilson lines $V$. This cross section is the amplitude diagram in Fig.~\ref{fig:DDIS&dipole}b squared.

\section{Setup}
We use an MV model~\cite{McLerran:1993ni} initial condition for the Wilson lines of the target nucleus on a transverse lattice. The Wilson lines are evolved using the JIMWLK equation. The initial geometry of the target Wilson lines is given by the IP-Glasma model~\cite{Schenke:2012wb,ipglasma_jimwlk_code} and the geometry evolution is given by the JIMWLK equation. The initial condition for the JIMWLK evolution has been fitted to exclusive $J/\psi$ production data~\cite{Mantysaari:2022sux}. The setup contains no free parameters aside from the quark masses.

The JIMWLK equation is implemented in the IP-Glasma code~\cite{ipglasma_jimwlk_code} using its Langevin formulation. The evolution step is
\begin{equation}\label{eq:jtimestepsymm}
V_\xt(y + \ud y) = 
\exp\left\{-i\frac{\sqrt{\as \ud y }}{\pi}\int_\zt 
  \Kt_{\xt-\zt} \cdot ( V_\zt \xit_\zt V^\dag_\zt) \right\}
\\
\times
V_\xt 
\exp\left\{i\frac{\sqrt{\as \dd{y} }}{\pi}\int_\zt 
  \Kt_{\xt-\zt} \cdot \xit_\zt \right\},
\end{equation}
where the evolution rapidity is $y=\ln(1/\xpom)$, $\alpha_s$ is the strong coupling constant and $\xit_\mathbf{z}=\xit_\mathbf{z}^a t^a$ is a local random Gaussian noise. The JIMWLK kernel is $\Kt_{\xt} = m |\xt|  K_1(m|\xt|) \frac{\xt}{\xt^2}$, where $m$ is a regulating mass, and $K_1$ is a modified Bessel function of the second kind. For a more detailed description, see eg. Ref.~\cite{Lappi:2012vw}.

The general form of the diffractive $q\bar q$ production cross section \eqref{eq:gammaAxsection} is discretized on the two dimensional transverse lattice
\begin{equation}
    \label{eq:numerical-integral}
    \begin{aligned}
        &\frac{\mathrm{d}\sigma^{\mathrm{D}}_{\lambda,q\bar q}}{\mathrm{d} M^2_X \mathrm{d}|t|}= \frac{N_{\mathrm{c}}}{4\pi}\frac{1}{4(2\pi)^3} 
        \sum_f e_f^2 \sum_{z_i=z_{\text{min}}}^{1/2} \Delta z 
        \sum_{\theta_{\mathbf{l}\mathbf{\Delta},i}=0}^{2\pi} \Delta\theta_{\mathbf{l}\mathbf{\Delta}}\sum_{h,\bar h = \pm1/2}  
        \Big| \Big\langle 
        \sum_{\mathbf{x}_\perp, \mathbf{y}_\perp}
        a^4 {\psi}_{\gamma^*_{\lambda}\to q\bar q} \\
        & \times \Big[ 1-\frac{1}{N_{\mathrm{c}}} \Tr[ V(\mathbf{x}_\perp) V^\dagger(\mathbf{y}_\perp)] \Big]
        e^{i \big( \mathbf{l} \cdot (\mathbf{x}_{\perp} - \mathbf{y}_\perp) + \mathbf{\Delta} \cdot (z_i \mathbf{x}_{\perp} + (1-z_i) \mathbf{y}_\perp) \big)}
        \Big\rangle\Big|^2,    
    \end{aligned}
\end{equation}
where $\lambda=\mathrm{T},\mathrm{L}$ is the polarization of the virtual photon, $e_f$ is the electric charge of quark flavor $f$, $z$ is the momentum fraction of the virtual photon carried by the quark, and ${z_\text{min}=1/2-\sqrt{1/4 - m_f^2/M_X^2}}$. The vectors $\mathbf{l}$ and $\mathbf{\Delta}$ are defined such that $\mathbf{l}$ is fixed to be along the x direction, their lengths are ${|\mathbf{l}| = \sqrt{z_i(1-z_i)M_X^2-m_f^2}}$ and ${|\mathbf{\Delta}|=\sqrt{|t|}}$ and their relative angle is $\theta_{\mathbf{l}\mathbf{\Delta}}$. The helicities of the quark and antiquark are $h$ and $\bar h$, and the coordinates $\mathbf{x}_\perp$ and $\mathbf{y}_\perp$ are the quark and antiquark positions. The angle brackets $\langle\cdot\rangle$ denote an ensemble average over Wilson line configurations with lattice constant $a$. The light cone wave function for the $q\bar q$ pair production is 
\begin{equation}
    {\psi}_{\gamma^*_{\mathrm{L}}\to q\bar q} = \frac{e}{2\pi}(z_i(1-z_i))^{3/2}2QK_0(\epsilon |\mathbf{r}|)\delta_{h,-\bar h}
\end{equation}
for the longitudinally polarized virtual photon, and
\begin{equation}
    {\psi}_{\gamma^*_{\mathrm{T}}\to q\bar q} = \frac{e}{2\pi}\sqrt{2}\sqrt{z_i(1-z_i)}\bigg(\sqrt{z_i^2+(1-z_i)^2}\epsilon \frac{\mathbf{r}\cdot \mathbf{\epsilon}^\sigma}{|\mathbf{r}|}K_1(\epsilon |\mathbf{r}|)\delta_{h,-\bar h}
    + 
    m_f K_0(\epsilon |\mathbf{r}|)\delta_{h,\bar h}\bigg)
\end{equation}
for the transversely polarized virtual photon. Here $K_0$ and $K_1$ are modified Bessel functions of the second kind, ${\epsilon = \sqrt{z_i(1-z_i)Q^2 + m_f^2}}$, ${\mathbf{r} = \mathbf{x}_\perp-\mathbf{y}_\perp}$ and $m_f$ is the mass of quark flavor $f$. For this case equation \eqref{eq:numerical-integral} is averaged over the two T polarizations $\mathbf{\epsilon}^\sigma$, where we use $\sum_\sigma \mathbf{\epsilon}^\sigma_i \mathbf{\epsilon}^\sigma_j = \delta_{i,j}$.

\section{Results}

\begin{figure}
    \centering
    \includegraphics[width=0.6\linewidth]{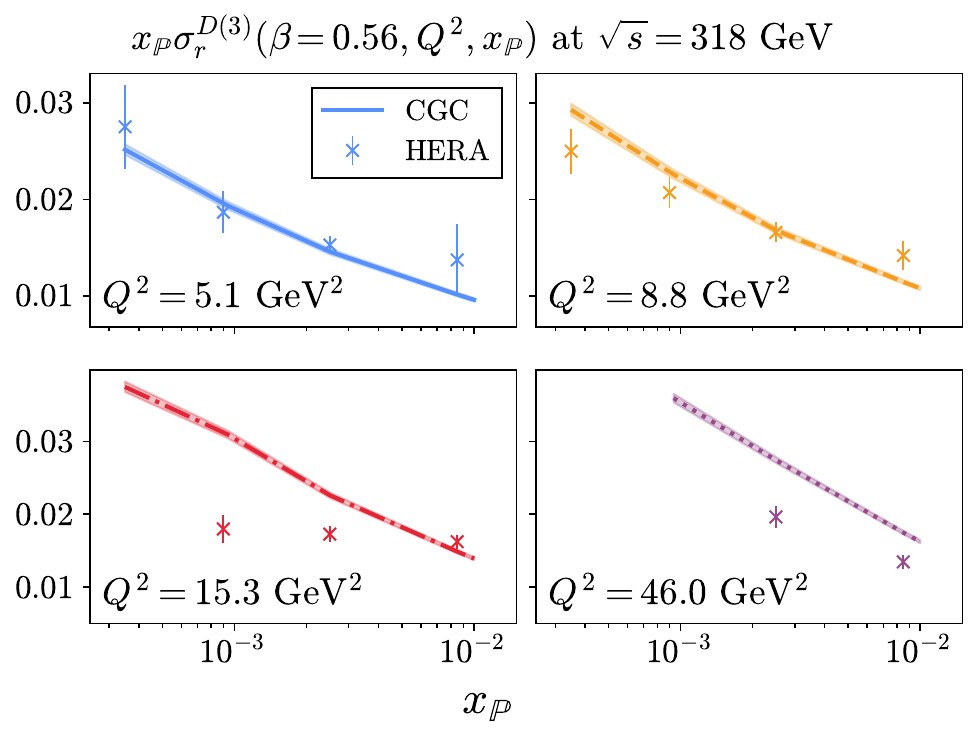}
    \caption{Proton reduced cross section $\xpom\sigma_r^{D(3)}$ at $\sqrt{s} = 318$ $\text{GeV}^2$ and four values of $Q^2$ as a function of $\xpom$ compared to HERA data~\cite{H1:2012xlc}. The $\xpom$ dependence is given by the JIMWLK evolution. The bands depict statistical errors.}
    \label{fig:p-F2D3-xpom}
\end{figure}

Our results for the proton reduced cross section are shown in Fig.~\ref{fig:p-F2D3-xpom} along with the combined data from HERA~\cite{H1:2012xlc}. The $\xpom$ dependence from the JIMWLK evolution is compatible with the data, but the $Q^2$ evolution of our results is faster. Similar systematics has been observed in previous calculations for inclusive $F_2$ in the dipole picture~\cite{Lappi:2013zma}. The good agreement at low $Q^2$ makes this comparison a successful benchmark.

\begin{figure}[t]
    \centering
    \includegraphics[width=0.6\linewidth]{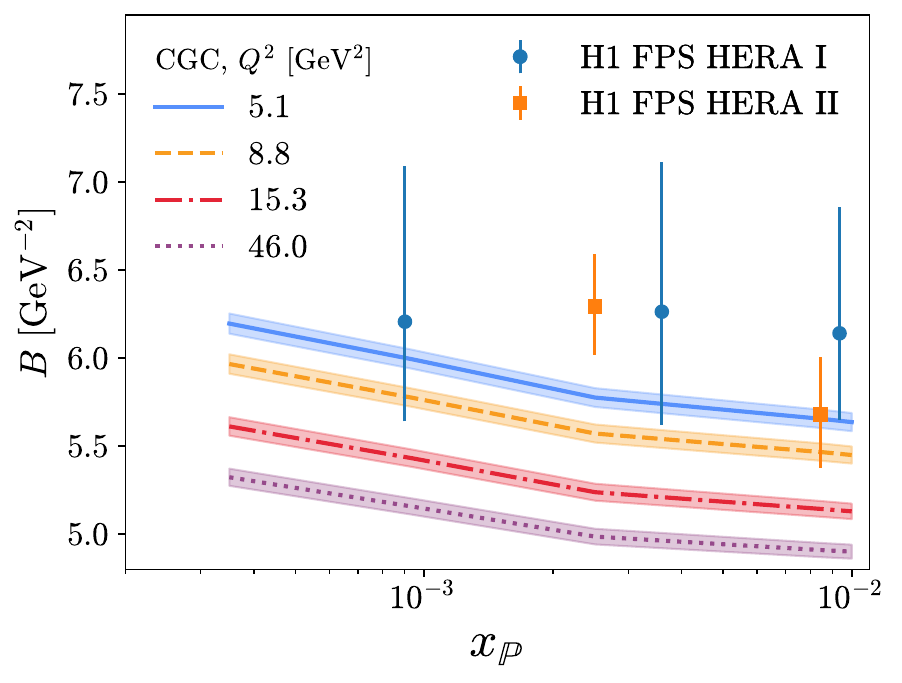}
    \caption{Proton slope parameter $B$ as a function of $\xpom$ at four values of $Q^2$ and corresponding data from H1 FPS at $Q^2 \in [5,10]$ $\mathrm{GeV}^{2}$ (HERA I~\cite{H1:2006uea}) and $Q^2 \in [4,110]$ $\mathrm{GeV}^{2}$ (HERA II~\cite{Aaron:2010aa}). Our $\xpom$ dependence is given by the JIMWLK evolution and the bands depict statistical errors.} 
    \label{fig:p-B-t}
\end{figure}

From the $t$ spectrum of the cross section we extract the proton slope parameter $B$. The $t$ dependence is parametrized as an exponential
\begin{equation}
    \frac{\text d \sigma_{\lambda,q\bar q}^{\mathrm{D}}}{\text d |t|} \sim e^{-B|t|},
\end{equation}
and the parameter $B$ is taken from a fit to the region $|t|<0.5$ $\text{GeV}^2$. The results for $B$ are shown in Fig.~\ref{fig:p-B-t}. Our results are in the same ballpark as the H1 data~\cite{H1:2006uea, Aaron:2010aa}. The results show an increase of $B$ towards smaller $\xpom$ that can be interpreted as an increase in the transverse size of the target color fields due to the radiation of gluons. An increase in $B$ also happens at smaller $Q^2$ where the photon typically fluctuates into larger $q\bar q$ dipoles. The H1 data has no measurable dependence on $Q^2$~\cite{Aaron:2010aa}.

\begin{figure}[h]
  \begin{tabular}{ll}
  \begin{minipage}[t]{0.45\textwidth} 
  \centering
    \includegraphics[width=\linewidth]{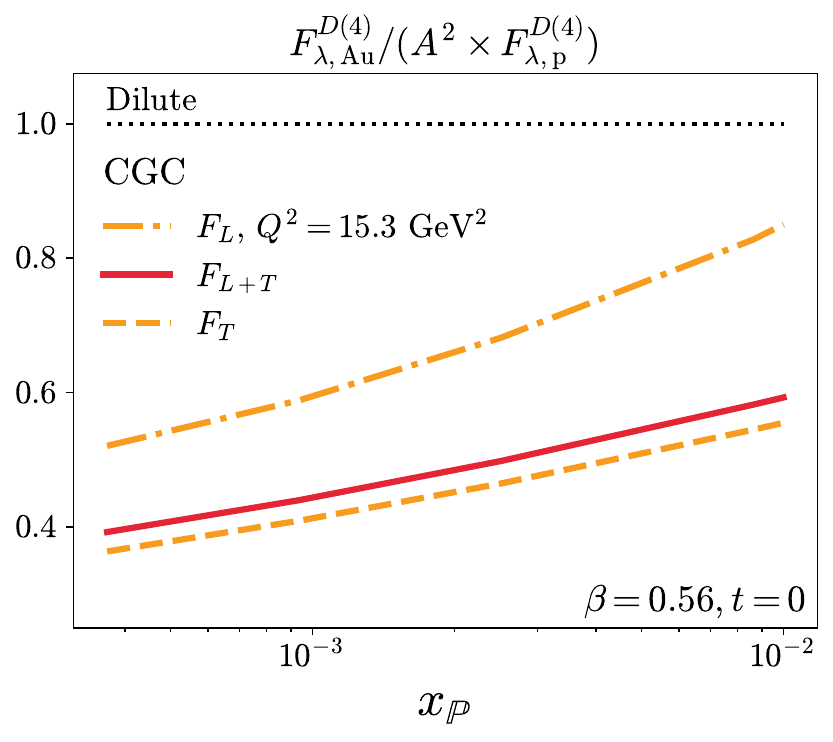}\\
     \textbf{(a)}
    \end{minipage}
    &
  \begin{minipage}[t]{0.45\textwidth} 
  \centering
    \includegraphics[width=\linewidth]{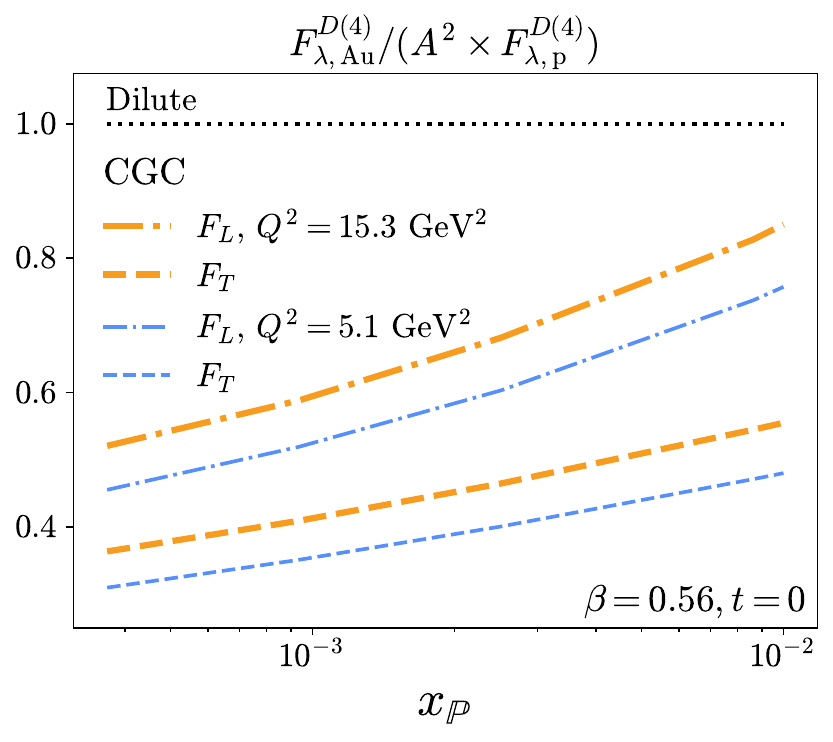}\\
    \textbf{(b)}
    \end{minipage}
  \end{tabular}
   \caption{\textbf{(a)} Nuclear modification factors for the diffractive structure functions of the gold nucleus $F_{\lambda,Au}^{D(4)}$ normalized by the square of the mass number $A=197$ computed at $\beta = 0.56$, $Q^2=15.3$ $\text{GeV}^2$ and $t=0$ with $\xpom$ dependence given by the JIMWLK evolution. The dilute limit (no saturation) is drawn in a dotted line. \textbf{(b)} Nuclear modification factors computed at two values of $Q^2$.}
    \label{fig:A-F2D3-t0}
\end{figure}

The main result of this work is the nuclear modification for the gold nucleus $F_{\lambda,\text{Au}}^{\mathrm{D}(4)}/(A^2\times F_{\lambda,\mathrm{p}}^{\mathrm{D}(4)})$, which is shown in Fig.~\ref{fig:A-F2D3-t0} at $t=0$ for one value of $Q^2$ and in Fig.~\ref{fig:A-F2D3-t0} for a comparison between two values of $Q^2$. The normalization $A^2$ is chosen such that this nuclear modification factor equals unity at $t=0$ in the dilute limit where saturation effects are absent. The results show a strong nuclear suppression, strongest for the transverse structure function which dominates the cross section in these kinematics. The larger suppression for the transverse case results from the transverse photon typically splitting into larger quark antiquark dipoles, which leads to non-linear multiple scattering. The suppression is seen to be stronger for smaller values of $Q^2$ in the latter figure for the same reason.

\section{Summary}

We have a setup for computing diffractive structure functions that is constrained by exclusive $J/\psi$ production data. We use the IP-Glasma model~\cite{Schenke:2012wb,ipglasma_jimwlk_code}, and the $\xpom$ evolution of the structure functions is given by the JIMWLK evolution. We achieve a good description of the HERA reduced diffractive structure function data for the proton. From the $t$ spectrum of the cross section we predict the slope parameter B and its $\xpom$  and $Q^2$ evolution showcasing the change in the geometry of the target-probe system in different kinematics. Our main result is the prediction of a strong nuclear suppression for the diffractive cross section in EIC kinematics for the gold nucleus. This same setup can be used to also compute diffractive-to-total cross section ratios, like the $eA$-to-$ep$ double ratio $[(\text d \sigma_{\text{diff}}/\text d M_X^2)/\text d\sigma_{\text{tot}}]_{eA}/(\text d \sigma_{\text{diff}}/\text d M_X^2)/\text d\sigma_{\text{tot}}]_{ep}$ presented as a key measurement for the EIC~\cite{Accardi:2012qut}. The predictions from this work can be compared to these future measurements.

\section*{Acknowledgments}

This work has been supported by the Research Council of Finland, the Centre of Excellence in Quark Matter (projects 346324 and 364191) and by the European Research Council (ERC, grant agreements No. ERC-2023-101123801 GlueSatLight and No. ERC-2018-ADG-835105 YoctoLHC). PR also acknowledges the support of the Vilho, Yrjö and Kalle Väisälä Foundation. Computing capacity from CSC--IT Center for Science in Finland and the Finnish Grid and Cloud Infrastructure (persistent identifier urn:nbn:fi:
research-infras-2016072533) were used in this work. The content of this article does not reflect the official opinion of the European Union and responsibility for the information and views expressed therein lies entirely with the authors.

\bibliography{refs}
\bibliographystyle{JHEP-2modlong}


\end{document}